# Controlled Microdroplet Transport in an Atmospheric Pressure Microplasma

P.D. Maguire[1,a], C.M.O. Mahony[1], C.P. Kelsey[1], A. Bingham[2], E.P. Montgomery[3], E.D. Bennet[4], H.E. Potts[4], D. Rutherford[1], D.A. McDowell[1], D.A. Diver[4], D. Mariotti[1]

[1]NIBEC, University of Ulster, BT37 0QB, N. Ireland
[2]Andor Technology, Belfast BT12 7AL,
[3]University of Bath, Bath, BA2 7AY
[4]SUPA, School of Physics and Astronomy, University of Glasgow, Glasgow G12 8QQ, UK

We report the controlled injection of near-isolated micron-sized liquid droplets into a low temperature He-Ne steady-state rf plasma at atmospheric pressure. The $H_2O$ droplet stream is constrained within a 2 mm diameter quartz tube. Imaging at the tube exit indicates a log-normal droplet size distribution with an initial count mean diameter of 15 μm falling to 13 μm with plasma exposure. The radial velocity profile is approximately parabolic indicating near laminar flow conditions with the majority of droplets travelling at > 75% of the local gas speed and having a plasma transit time of < 100 μs. The maximum gas temperature, determined from nitrogen spectral lines, was below 400 K and the observed droplet size reduction implies additional factors beyond standard evaporation, including charge and surface chemistry effects. The successful demonstration of controlled microdroplet streams opens up possibilities for gas-phase microreactors and remote delivery of active species for plasma medicine. © *American Institute of Physics.* [DOI: unknown]

[1]The interaction between non-thermal plasmas and liquids has recently encouraged intensive research into fundamental processes and emerging technological applications including plasma medicine, nanomaterials synthesis and decontamination among others [1]. Stability and control of the plasma in contact with liquid is a key challenge in this highly non-equilibrium state. Plasma configurations involve either liquid electrodes or discharges formed directly in liquids or within bubbles in liquids. Recently, water sprays or jets have been injected into the plasma zone to enhance the surface area of the liquid and improve energy efficiency for bacterial inactivation [2] or to enable continuous nanoparticle production [3]. However, injecting and controlling isolated droplets within a non-thermal plasma would offer new opportunities for fundamental studies and modelling of the complex electron – liquid non-equilibrium chemistry [4]. It also opens up the prospects of new technological applications such as gas-phase microreactors, which while analogous to their microfluidic counterparts with multiphase emulsion flows [5,6,7] plasma-liquid systems offer additional chemistries, materials and a route to large scale implementation [8,3]. In plasma medicine plasma activated droplets will also enhance the scope for remote delivery of therapeutic agents to biological tissue and organs [9].

To maintain the stability and low temperature of the plasma, a small diameter (2 mm) capillary is required while rf excitation allows steady-state non-thermal equilibrium operation. Injecting divergent droplet sprays into a narrow orifice is required while maintaining droplet integrity and matching gas flows for stable plasma operation. In this paper we report exit droplet size and velocity distributions with and without plasma exposure along with plasma electrical characteristics and gas temperature measurements from optical emission spectroscopy.

A continuous stream of near mono-sized droplets was

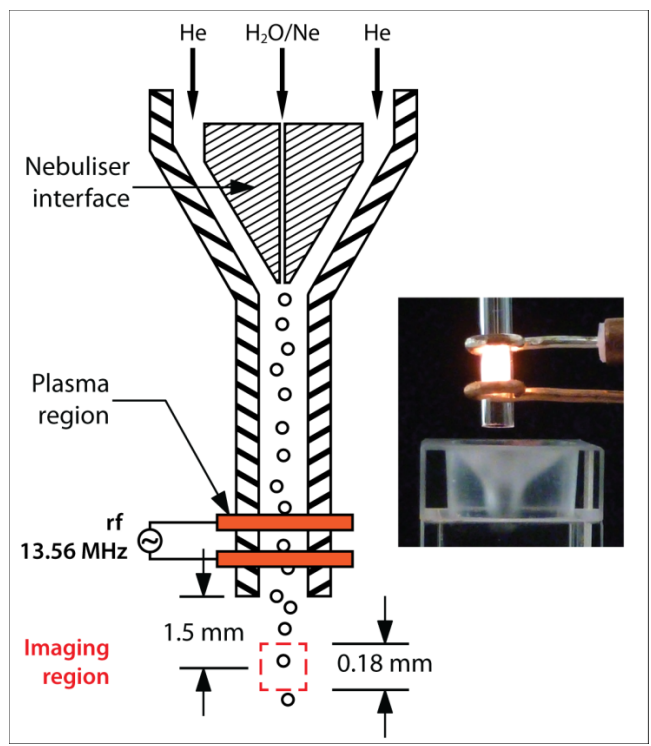

FIG. 1. Schematic of plasma set up consisting of ring electrodes around 2 mm diameter quartz tube and nebuliser interface. The imaging region is 1.5 mm from tube exit. Inset photo shows gas/droplet stream in liquid-filled cuvette.

obtained from a Burgener Mira Mist X-175 nebuliser driven by a neon gas flows up to 1.0 slm via a mass flow controller. The liquid ($H_2O$) was supplied by a syringe pump at 3 – 16 μl/min. The divergent (20°) aerosol output of the nebuliser was interfaced to a quartz tube with a 2 mm inner diameter via a custom-designed acrylic manifold which focussed the aerosol/Ne gas flow using a concentric helium gas curtain with controlled flows between 1 – 4 slm. The aerosol droplet transport efficiency is optimised by varying the water, neon and helium gas flows. The He/Ne gas stream

---

[1a] Author to whom correspondence should be addressed: pd.maguire@ulster.ac.uk

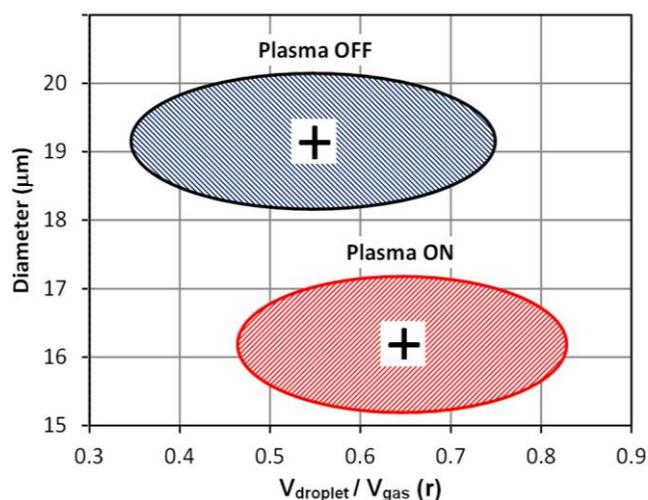

FIG. 2. Droplet diameter and velocity normalised to the radial gas velocity, showing arithmetic mean values with and without plasma. The shaded areas represent 1 standard deviation for both parameters

transported the droplets along the tube to the plasma region formed by concentric ring electrodes at the outer wall and driven by rf power from 80 W to 120 W at 13.56 MHz. The length of the plasma region between electrodes was ~2 mm with a further ~1 mm to tube exit, Figure (1). Low resolution images show a parallel droplet spray at the tube outlet, indicating persistent laminar flow beyond the capillary, with a slight reduction in spray diameter when the plasma is turned on. Droplet velocity, position and diameter (D) after exiting the plasma were obtained from high resolution images of area 1805 x 1805 μm centred at 1500 μm and 4000 μm from the end of the tube. A Questar QM-100 long focal length microscope was used to image the droplets onto an Andor IStar CCD camera with exposures of 10 μs or 20 μs. The object plane coincided with the central axis of the aerosol flux and quartz tube and the depth of field was 150 μm. Velocity and diameter were determined from streak length and width respectively.

A simple double Langmuir probe consisting of a pair of 0.3 mm diameter wires (0.6mm apart) was placed in the plasma afterglow at the tube exit to measure electron temperature. An estimate of electron temperature ($T_e$) of $5 \pm 2$ eV was obtained from measured current for a plasma operated at 100W with and without aerosol. An Impedans Octiv IV probe was used to measure the RF current, voltage and phase on the incoming RF power line for a range of helium and neon flow rates. The quartz capillary, for impedance measurements only, had an internal diameter of 0.7 mm. Under lower gas flow (Ne/He 0.7/1.0 slm) and power conditions (40 W – 70 W) the time-averaged electron density ($N_e$) was estimated, from the plasma resistance, to vary from $3 \times 10^{12}$ cm$^{-3}$ to $7 \times 10^{12}$ cm$^{-3}$ without and with aerosol respectively. $N_e$ was observed to increase with power and neon concentration. Optical emission spectra in the wavelength range 300 – 390 nm were obtained using an Ocean Optics HR4000 spectrometer at a plasma power of 120 W. Gas temperature was estimated from spectral fits to synthetic spectra ($N_2$ and OH lines) generated by Specair software. For a He plasma with 0.1%$N_2$ (i.e. 0% Ne and without aerosol), average gas temperature estimates were ~335K for flow rates > 2 slm but this increased to ~600K as flow is reduced to 0.15 slm. This temperature – flow dependence is similar to that previously reported for dc microdischarges [10]. For He-Ne gas flows (ratio 3.5:0.5), with and without aerosol (10 μL min$^{-1}$), the integration time was 30 seconds per spectrum and 95 background corrected spectra were taken and averaged.

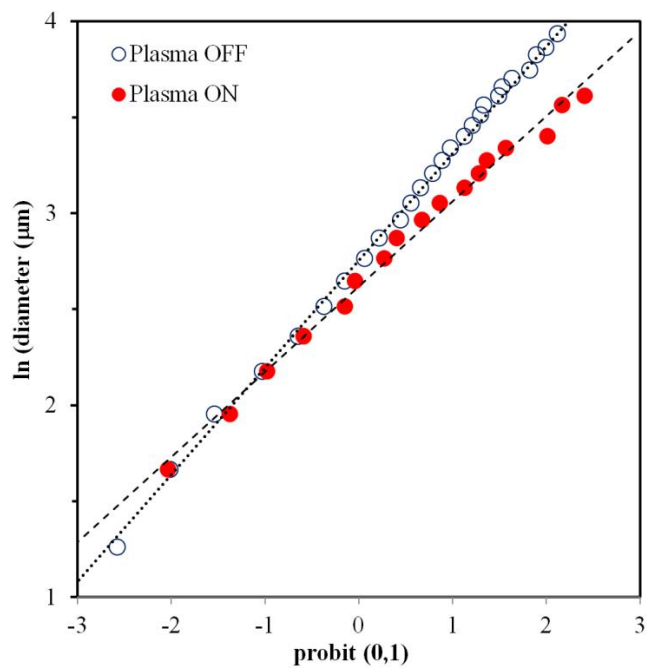

FIG. 3. Log-probability plot of ln(D) in μm against the inverse standard normal distribution function (probit) for droplet size distributions with and without plasma. The dotted lines in each case are least-square linear fits with $R^2$ values $> 0.99$.

From individual droplet size and velocity measurements at the tube exit, for an average gas velocity of 32 ms$^{-1}$, a mean diameter reduction of 3 μm is observed, from 19 μm to 16 μm, along with ~20% increase in average velocity, figure (2), for droplets transported through the plasma. The measured velocity increase may be due to evaporation of slower droplets to below the imaging threshold. In the log-probability plot, figure (3), ln(D) is plotted against the inverse normal cumulative distribution function (with zero mean and unit std. deviation) [11]. For plasma off and on conditions, good straight-lit fits were observed indicating log-normal aerosol diameter distributions with count median diameters (CMD) of 15.7 μm and 13.7 μm respectively and geometric standard deviations (GSD) of ~1.6. The average transport time was 93 μs (plasma on) – 113 μs (plasma off) which is >1000 rf periods. The input size distribution is unknown; however for similar nebulisers the, albeit limited, literature reports [12] log-normal distributions with CMD values ~10 – 15 μm, similar to that observed here at the exit, and an indication therefore of non-turbulent conditions at the tube entrance.

The axial velocity distributions as a function of radius, taken at 1.5 mm from tube exit, are shown in figure (4a) normalised to the average gas speed (32 ms$^{-1}$) and fall within a parabolic envelope, which indicates laminar gas and droplet flow within the tube and at its exit. A similar parabolic characteristic was observed for droplets at 4 mm from the exit and at lower gas speeds, indicating that little turbulence is induced at the tube interface with the ambient air. This is also observed with low resolution images of the exit spray. We can be confident therefore that most droplets within optical depth of field are imaged. Droplet axial velocity is ~75% of the local gas speed, figure (4b) except near the tube walls, where it is reduced considerably, indicating a possible interaction with the quartz surface or plasma sheath. After transport through the plasma, the near-wall region appears to be more depleted of droplets, which is also observed in exit spray images. Modification of the drag forces near the wall can be expected but estimates indicate it is not significant for distances > D from the wall, while the droplet loss fraction through diffusion to the wall, assuming laminar flow, is estimated to be much less than $10^{-3}$.

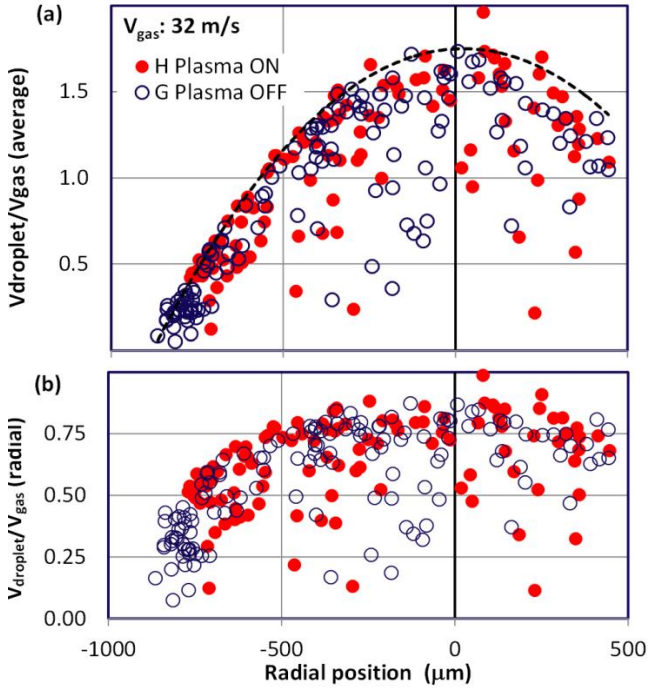

FIG. 4. Axial droplet velocities versus radial position for plasma (filled) and aerosol only (open) with respect to (a) the average gas velocity of 32 ms$^{-1}$ and (b) the axial droplet velocity normalised to gas velocity against radial position. The quartz wall is at position -1000.

The particle Reynolds number (Re) depends on diameter and the velocity difference between it and the fluid. For droplets, of size ≤ CMD and travelling at >75% of the gas speed, Re < 1 and the flow around the droplet is approximately laminar, an important factor in the convective element of evaporation and droplet mixing in flight. For larger droplets therefore, the local environment near the surface may have a degree of turbulence. The velocity relaxation time for Ne-carried droplets introduced into the He gas stream varies from ~100 μs to milli-seconds, depending on diameter and given the radial gas velocity profile, droplets will not fully equilibrate with the gas flow over the distance from nebuliser output to the imaging position at the tube exit. This is particularly so for larger particles and hence collisions and coagulation with advancing smaller particles is a possibility. However, for ~25 x $10^3$ droplets/sec at an average gas speed of 32 ms$^{-1}$, there are only 10 – 100 droplets in the tube at any instant and hence significant collision-induced coagulation is not thought likely.

Figure (5) shows spectra with and without aerosol. Line identification of the N2 molecular bands [13 & 14] and OH (A-X)

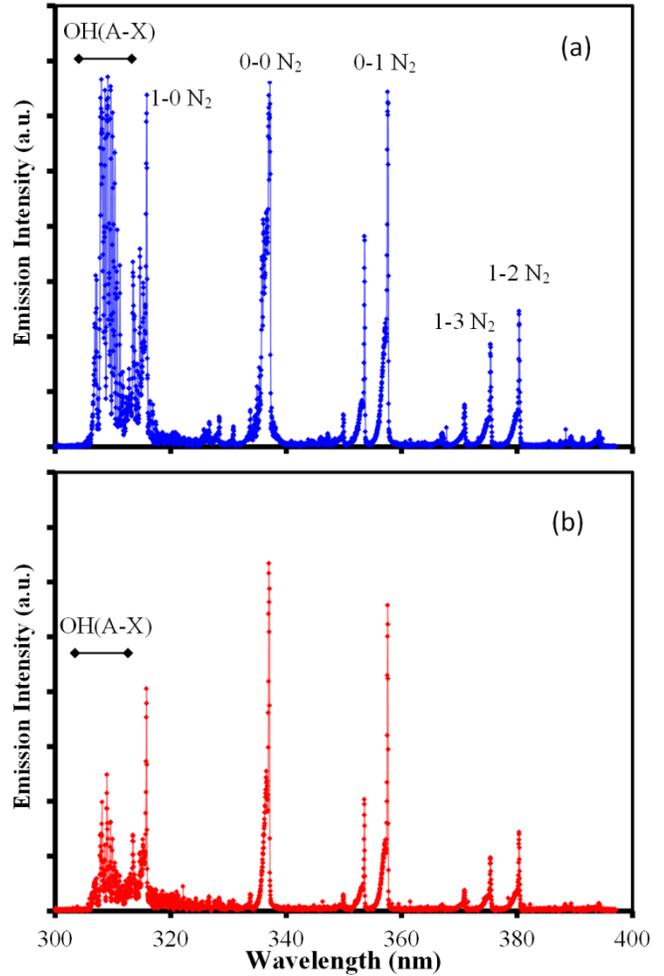

FIG. 5. Optical emission spectra indicating N$_2$ and OH(A-X) emission lines for (a) plasma – no aerosol and (b) plasma with aerosol.

lines [13] are given and with the aerosol flow, the reduction in OH line intensities is clearly visible. For the case with aerosol N$_2$ rotational temperature ($T_{rot}$) falls within the range 280-350 K with the best fits obtained at 330K, while with the aerosol the corresponding temperature range is 330-400K. The gas temperatures remained constant in time. Neon containing atmospheric pressure plasmas are rare in the literature [1]. However, with the exception of argon containing plasmas, it is unlikely that the rotational population determined from N$_2$ lines is not in equilibrium with the gas and therefore the rotational temperature is assumed to be a good approximation of the background gas temperature. Verreycken et al. however [15] found a deviation between $T_{rot}$(OH) and gas temperature obtained from Rayleigh scattering for a dc plasma with a water electrode. Montret et al. [16] found, in a humid argon DBD plasma, the disappearance of OH(A – X) fluorescence to be dominated by collisional de-excitation with water molecules at high vapour content and the intensity – vapour characteristic peaks at 300ppm H$_2$O in argon. Assuming that in He – Ne mixtures, a similar peaked OH fluorescence characteristic exists then the observed reduction in OH intensity implies that humidity of the gas stream increases significantly in the presence of aerosol droplets. For a droplet rate of 25 x $10^3$ s$^{-1}$, there is an average of 2 droplets in the plasma at any one time and the evaporation of these droplets by 2 μm from the measured CMD values contributes about 200ppm water vapour into the plasma volume. This would appear to be insufficient to explain the OH fluorescence quenching unless the collisional de-excitation is enhanced by the presence of He – Ne over that observed with Ar.

The evaporation rate of sessile water droplets typically follows a D-squared law ($D^2 = D_o^2 - ct$) where $c$ is an empirical evaporation constant [11]. We have fitted the distributions with and without plasma using 10 velocity bins to estimate times of flight and from this determined a $c$ value of $5 \times 10^{-7}$ m$^2$s$^{-1}$ that represents the enhanced evaporation due to plasma exposure. Heinisch et al. [17] measured diameter and temperature of charged water droplets in an electrodynamic trap and determined an evaporation rate of $< 1.5 \times 10^{-10}$ m$^2$s$^{-1}$ for stagnant humid N$_2$ and $1.5 \times 10^{-9}$ m$^2$s$^{-1}$ for flowing dry N$_2$. These values are 2 – 3 orders of magnitude lower than observed in our case. Assuming an effective plasma heating zone that extends to beyond the imaging region makes little difference ($c = 3 \times 10^{-7}$ m$^2$s$^{-1}$). Benson et al. [18] modelled droplet heating in high temperature Ar (3000 K) at atmospheric pressure and estimated an evaporation rate of $5 \times 10^{-7}$ m$^2$s$^{-1}$ and wet bulb temperature of 365 K, i.e. below boiling point. It is clear therefore, given the estimated gas temperatures from OES of < 400 K and the very short transit times that the observed reduction in diameter cannot be simply due to evaporation nor diffusion losses to the wall. Interaction with the plasma sheath and/or the evaporation of more volatile species formed at the droplet surface due to electron bombardment [4] may offer possibilities and are worthy of further investigation. Using capacitance values determined from plasma impedance measurements, we estimate that the sheath almost doubles in thickness on the introduction of the aerosol and the electron density increases. While this is a rudimentary sheath model, it does highlight the impact of the aerosol on the plasma characteristics that may be similar to the case of complex (dusty) plasmas, despite the limited number of droplets. In dusty plasmas the droplet charge, which depends mainly on diameter and electron temperature, is the most significant parameter. We have shown, for small (< 100 nm) solid nanoparticles in a similar plasma, that collision induced heating is significant, leading to surface temperatures > 700 K for gas temperatures < 400 K [19]. The impact on relatively large microparticles in a highly collisional regime is not well understood although reduction in ion flux due to ion-neutral collisions is possible and this may reduce the droplet charge and heating [20]. However fragmentation of the droplet surface, due to charge-induced instabilities, into sub-micron sized droplets may well have more significant impact on the plasma and on fragment heating.

Controlled injection of near-isolated liquid microdroplets into a low-temperature atmospheric pressure plasma has been achieved. Low Reynolds number and laminar flow conditions pertain. Given the short transport times (~100 μs) and gas temperatures < 400K, the average reduction in count median diameter would require an anomalously high evaporation rate. The estimated increase in humidity due to droplet size reduction is around 200 ppm, although quenching of the OH fluorescence, compared to the aerosol-free plasma, implies a much greater humidity value. Volatile species formation at the surface due to electron bombardment may lead to enhanced evaporation, while microdroplet charging and impact on plasma conditions are also important factors worth further investigation.

This work was supported by the Engineering and Physical Sciences Research Council (EPSRC: EP/K006088/1, EP/K006142 & EP/K022237/1), Royal Society International Exchange Scheme (IE120884), the Leverhulme International Network (IN-2012-136) and NATO SPS(EAP.SFPP 984555).